\documentclass[a4paper,fleqn,usenatbib]{mnras}
\pdfoutput=1

\usepackage[T1]{fontenc}
\usepackage{ae,aecompl}

\usepackage{graphicx}	
\usepackage{amsmath}	
\usepackage{amssymb}	
\usepackage{caption}

\newcommand{\angstrom}{\mbox{\normalfont\AA}}

\title[Composite Seyfert 1 spectrum]{Seyfert 1 Composite Spectrum using SDSS Legacy Survey Data}

\author[N. Pol et al.]{
	Nihan Pol$^{1}$\thanks{E-mail: nspol@mix.wvu.edu} and 
	Yogesh Wadadekar,$^{2}$ \thanks{E-mail: yogesh@ncra.tifr.res.in}
	\\
	$^{1}$Department of Physics and Astronomy, West Virginia University, 135 Willey Street, Morgantown, WV 26506, United States \\
	$^{2}$National Centre for Radio Astrophysics, TIFR, Post Bag 3, Ganeshkhind, Pune 411007, India\\
}

\date{Accepted XXX. Received YYY; in original form ZZZ}

\pubyear{2016}

\begin{document}
	\label{firstpage}
	\pagerange{\pageref{firstpage}--\pageref{lastpage}}
	\maketitle
	
	\begin{abstract}
		
		We present a rest-frame composite spectrum for Seyfert 1
		galaxies using spectra obtained from the DR12 release of the Sloan
		Digital Sky Survey (SDSS). The spectrum is constructed by combining
		data from a total of 10,112 galaxies, spanning a redshift range of $0$
		to $0.793$. We produce an electronic table of the median and geometric
		mean composite Seyfert 1 spectrum. We measure the spectral index of
		the composite spectrum, and compare it with that of the composite
		quasar spectrum. We also measure the flux and width of the strong
		emission lines present in the composite spectrum. We compare the
		entire spectrum with the quasar spectrum in the context of the AGN
		unification model. The two composite spectra match extremely well in
		the blue part of the spectrum, while there is an offset in flux in the
		red portion of the spectrum.
	\end{abstract}
	
	\begin{keywords}
		galaxies: Seyfert  -- galaxies: active -- techniques: spectroscopic -- surveys
	\end{keywords}
	
	
	
	\section{Introduction}
	
	Seyfert galaxies are one of the most common types of Active
	Galactic Nuclei (AGN) in the nearby Universe. Amongst these
	galaxies, there are two major subclasses called Seyfert Type 1 and
	Seyfert Type 2 which are distinguished by their optical spectra. The spectra
	of Type I Seyfert galaxies show broad lines that include both
	allowed lines, and narrower forbidden lines. In Type I Seyfert
	galaxies, we are able to observe the central compact source more or
	less directly, therefore sampling the high velocity clouds that
	produce the broad emission lines from a region close to the
	supermassive black hole at the center of the galaxy. Seyfert nuclei
	are typically hosted by spiral galaxies. They are believed to be the
	low luminosity and low redshift counterparts of Type 1 quasars seen at
	higher redshifts.
	
	The optical spectra of AGN, be they quasars, Seyfert 1s or Seyfert2s,
	can be thought of as the sum of a featureless continuum and emission
	lines, which may be broad or narrow.  Such a generic similarity
	between the spectra of different types of AGN is one of the
	drivers behind the unified model for AGN (\citet{unifiedagn}, \citet{new_unified_agn}). The basic principle of the unified model is that the underlying
	physical scenario for all AGNs is intrinsically similar and AGN
	diversity is just a geometric effect caused by different orientation
	angles of the disk with respect to the observer. AGNs of the
	same subclass such as blazars, Seyfert 1s etc. have spectra with a
	number of common characteristics. Due to this, the individual spectra
	of specific subtypes of AGN can then be effectively combined
	into one single composite spectrum, which can be used to determine
	several median or average properties of the members of that AGN class.
	
	For quasars, a number of efforts have been described in the
	literature to obtain such a composite spectrum and to use it for various
	scientific investigations.  For instance, a composite spectrum was used to determine
	relative line strengths which in turn, was used to determine quasar
	structure \citep{Francis1991}. The composite spectrum was also used
	as a cross-correlation template for identification and redshift
	determination of quasars in SDSS data \citep{VandenBerk2000}. A composite
	spectrum with sufficiently high signal to noise ratio can be used to
	find hitherto undiscovered emission lines in the spectra of quasars
	(\citet{VandenBerk2001}, \citet{Francis1991}). In addition to this,
	they can also be used for precision measurement of emission line
	shifts relative to laboratory wavelengths (\citet{VandenBerk2001}),
	calculation of quasar colors for improved candidate selection
	(\citet{Richards2002}) and for calculating K-corrections used in
	evaluating the quasar luminosity function \citep{Kcorrection1} as well as for computing the mean colours of quasars at different redshifts
	\citep{White}.
	
	An interesting application of a composite spectrum can be seen in
	\citet{White}. In order to study the radio properties of quasars,
	\citet{White} make use of data from the VLA FIRST survey
	\citep{FIRST} in order to stack quasars from SDSS DR3 quasar catalog
	\citep{DR3}. Using these data, they stack the quasars in bins
	based on the parameter of interest, such as redshift, or optical
	magnitude (see \citet{White} for more details). In particular, to
	study the relationship between radio luminosity and absolute
	magnitude, they compute the redshift dependent K-corrections using the
	\citet{VandenBerk2001} composite quasar spectrum. If such a study for
	Seyfert 1 galaxies were to be undertaken, it would be necessary to
	have a composite spectrum for Seyfert 1 galaxies to compute the
	K-corrections in order to determine the radio luminosity-absolute
	magnitude relationship. This was one of the motivations behind this
	work.
	
	For quasars, there have been several previous works aimed towards generating such a
	composite spectrum. \citet{Francis1991} generated a composite
	quasar spectrum using 718 quasars from the Large Bright Quasar Survey
	\citep{LBQS}. Another composite spectrum was created by
	\citet{Zheng1997} from data obtained for 284 quasars using the Hubble
	Space Telescope (HST) Faint Object
	Spectrograph. \citet{Brotherton2001} also prepared a composite
	spectrum from 657 quasars in the FIRST Bright Quasar
	Survey. \citet{VandenBerk2001} generated a spectrum from 2204 spectra
	obtained from the commissioning phase of Sloan Digital Sky Survey
	(SDSS) \citep{VandenBerk2000}. This spectrum was thereafter used
	as a cross-correlation template throughout the SDSS project for
	classification of quasars, and determining their redshifts. The most
	recent construction of such a composite quasar spectrum based on SDSS data was done by
	\citet{Harris2016}, which contains observations of 102,150 quasars from
	the Baryon Oscillation Spectroscopic Survey \citep{BOSS} which is a
	part of SDSS III \citep{SDSS3}.
	
	In comparison, there have been relatively few attempts to create such
	a composite optical spectrum exclusively for Seyfert galaxies, which
	also belong to the AGN class, but are less luminous than their quasar
	counterparts. An analysis was done by \citet{comps11}, who analyzed
	spectra of 27 Seyfert 1 nuclei in the ultraviolet band, obtained from
	The International Ultraviolet Explorer \citep{IUE}. In this work,
	they binned the Seyfert 1 galaxies in three ranges of absolute
	luminosity, and then stacked the spectra in each of those bins to
	produce a composite Seyfert 1 spectrum. In addition to this, they also
	produced a Seyfert 1 spectrum of all the galaxies stacked together,
	irrespective of luminosity. They presented all these spectra in two
	wavelength windows, long (approximately $2000 \angstrom$ to $2900
	\angstrom$) and short (approximately $1200 \angstrom$ to $1900
	\angstrom$). \citet{comps11}, however, did not measure the spectral
	index for their sample of Seyfert 1 galaxies, with the study largely
	focused on examining emission features in the Seyfert 1 spectra. They
	compared their composite spectrum with a composite quasar spectrum
	prepared by \citet{oldqsospec} and found that the two spectra were
	identical, with all the emission features observed in quasars also
	observed in Seyfert 1s with similar equivalent widths.
	
	In this paper, we attempt to create a composite spectrum for Seyfert 1
	galaxies which will be useful to study their global properties, and
	compare them with the corresponding quasar properties, particularly
	those listed in \citet{VandenBerk2001}. Specifically, we wish to
	compare the continuum shape of Seyfert 1 galaxies to that of (Type 1)
	quasars, to test consistency with the AGN unification model. Thus,
	in a way, our work will be an implementation of the idea first
	developed by
	\citet{comps11}, but applied to optical wavelengths and with a much
	larger sample size. We also provide  electronic versions of several
	variants of our composite Seyfert 1 spectrum.
	
	Throughout this paper, we assume the $\Lambda CDM$ cosmology with $H_0
	= 70$ \ km \ s$^{-1}$ \ Mpc$^{-1}$, $\Omega_M = 0.3$, and
	$\Omega_{\Lambda} = 0.7$.
	
	In Section~\ref{sample}, we describe the sample used for construction
	of the composite Seyfert 1 spectrum. In Section~\ref{methodology}, we
	describe the methodology implemented for the creation of the spectrum,
	while Section~\ref{features} describes the continuum, emission and
	absorption features of the composite spectrum. Finally, in Section
	\ref{compare with quasar} we compare our results with those from
	\citet{VandenBerk2001}. Section~\ref{conclusion} contains the conclusion.
	
	\section{The Sample} \label{sample}
	
	\subsection{Seyfert 1 galaxies in SDSS}
	
	The candidates for the analysis presented in this paper were obtained
	from the most recent 13th edition of the long standing catalog of
	quasars and AGN published in \citet{VeronCetty2010}. This is a
	comprehensive catalog consisting of all known quasars and AGNs at
	the time of publication. The authors provide a dedicated table of
	AGNs, sub-classified as Seyfert 1, Seyfert 2, intermediate Seyferts,
	and LINERS (Table\_AGN in \citet{VeronCetty2010}). The table lists a
	total of 34,231 objects, of which 17,442 are Seyfert 1, 6024 are
	Seyfert 2, 878 are LINERs (classified as S3 in the table), and 9,887
	are unclassified. Because of the significantly larger sample, we chose
	to work with Seyfert 1 galaxies only.
	
	For the work presented here, we have selected only those objects which
	are neither classified as Seyfert 2s nor LINERs. We have also excluded
	unclassified objects, even though \citet{VeronCetty2010} suggest for
	them to be included as Seyfert 1 galaxies. This is due to the fact
	that these unclassified objects were originally classified as QSOs, but
	were fainter than $M_B = -22.25$ and were consequently added to the
	AGN table. Thus, to prevent undue contamination by possible QSO type
	sources, we have excluded them from our sample.
	
	\subsection{Obtaining the Spectra}
	
	For each of the Seyfert 1s in the Veron-Cetty catalog, we
	searched for the availability of an SDSS spectrum.  To do this, our
	sample was run through the SDSS CasJobs service in order to return the
	matches for these objects from the SDSS DR12 \citep{DR12}
	database. The identification parameters for these objects, viz. plate,
	MJD, and fiber ID were obtained from the SpecObjAll table in CasJobs.
	
	The requisite query returned a match with 12,429 galaxies, with a
	positional matching radius of $3''$. Of these 12,429 galaxies, 300 had
	a non-zero zWarning flag indicating some issue with redshift
	determination, and were discarded. We were then left with 12,129
	galaxies.
	
	The SDSS is a family of spectroscopic surveys comprising of Legacy
	(\citet{LEGACY}), BOSS (\citet{BOSS}), and SEGUE (\citet{SEGUE}) each
	with different strategies for targeting objects for spectroscopy, motivated by differences in their scientific goals. Of
	the 12,129 objects, 11,788 were drawn from the Legacy survey database,
	338 from the BOSS survey database, and the remaining 3 were from the
	SEGUE databases. Spectra for all these objects are available from the SDSS 3 Science Archive Server (SAS). We have chosen to work only with the objects from the Legacy survey because of homogeneity in target
	selection. Additionally, the fiber diameter also changed from 3 arcsec
	in Legacy to 2 arcsec in BOSS. The SEGUE survey being designed
	primarily for observation of stars, the few spectra of Seyfert 1s in
	it are chance detections. Besides, their number is very small (3 galaxies) and
	are also excluded from our analysis.
	
\begin{figure*}
		
		\includegraphics[width = \textwidth]{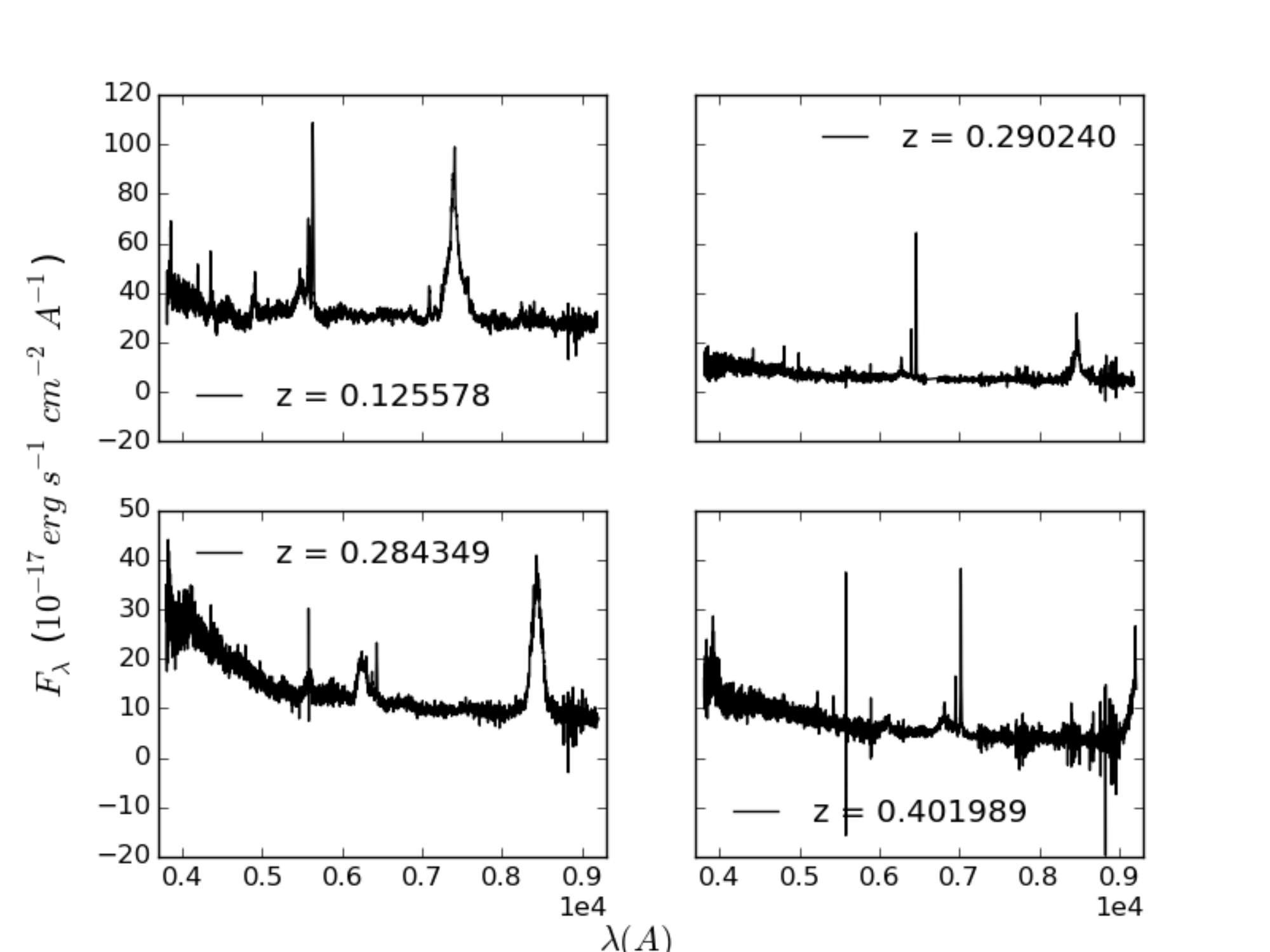}
		\caption{Some examples of the Seyfert 1 spectra that were used in preparing the composite spectrum. These spectra have been picked randomly from different redshift ranges.}
		\label{sample spectra}
		
	\end{figure*}

	These Legacy spectra have a wavelength coverage of $ 3800-9200
	\angstrom $, with spectral resolution in the range 1800-2200. A few of
	these spectra are shown in Figure~\ref{sample spectra}. The spectra
	shown were randomly selected from different redshift bins,  spanning as wide a redshift range as possible.
	
	In addition to the above caveats, SDSS also employs a spectroscopic
	classification system to automatically classify a given spectrum
	during data processing. This system has three classes of
	GALAXY, QSO, and STAR, with the GALAXY class having three sub-classes, STARFORMING, STARBURST, and AGN. The input spectrum is fit with either a galaxy, QSO, or stellar template model at different redshifts. The template which provides a fit with the least chi-squared value determines the class of the object. In the case of classification
	as a GALAXY spectrum, if the spectrum has detectable emission lines
	that are consistent with being a Seyfert galaxy or a LINER, in
	addition to satisfying the criteria:
	
	$\displaystyle log_{10} \left(\frac{OIII}{H\alpha}\right) > 0.7 - 1.2 \left(log_{10} \left(\frac{NII}{H\alpha}\right) + 0.4\right)$ then the spectrum is sub-categorized as an AGN\footnote{http://www.sdss.org/dr12/spectro/catalogs/}.
	
	To avoid contaminating our sample with objects whose SDSS spectra are
	not consistent with them being AGN, such objects were removed. Thus, we
	chose objects which fell under the categories of QSO or AGNs under
	GALAXY class, while rejecting all others with a different
	spectroscopic classification. Doing this leaves us with 10,162 Seyfert 1 galaxies.
	
	\begin{figure}
		\centering
		\includegraphics[width = \columnwidth]{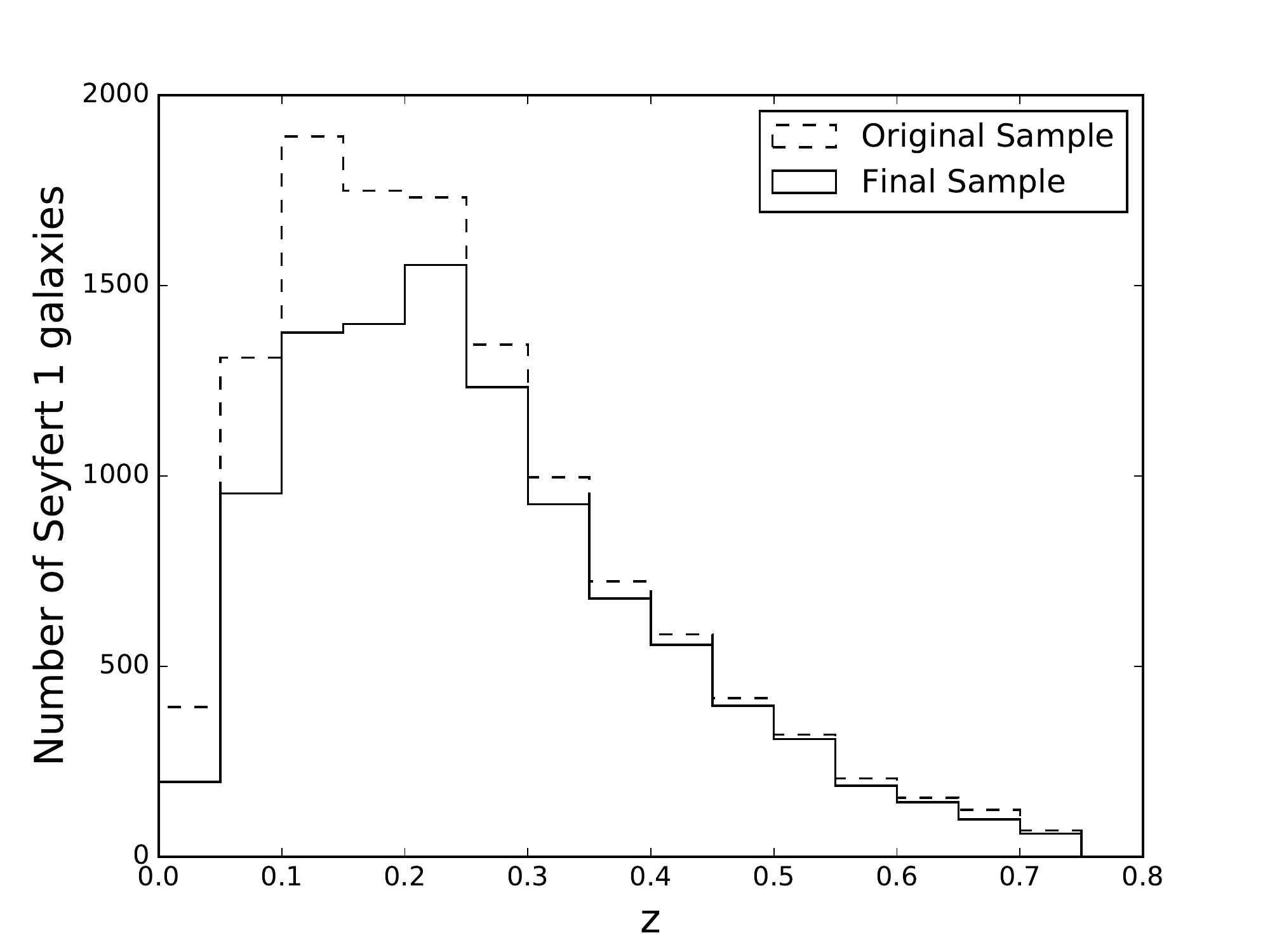}
		\caption{Histogram of number of galaxies per redshift bin. The number of galaxies beyond redshift of 0.8 is negligible, and hence they have been excluded from the sample.}
		\label{redshift hist}
	\end{figure}
	
	\begin{figure}
		\centering
		\includegraphics[width = \columnwidth]{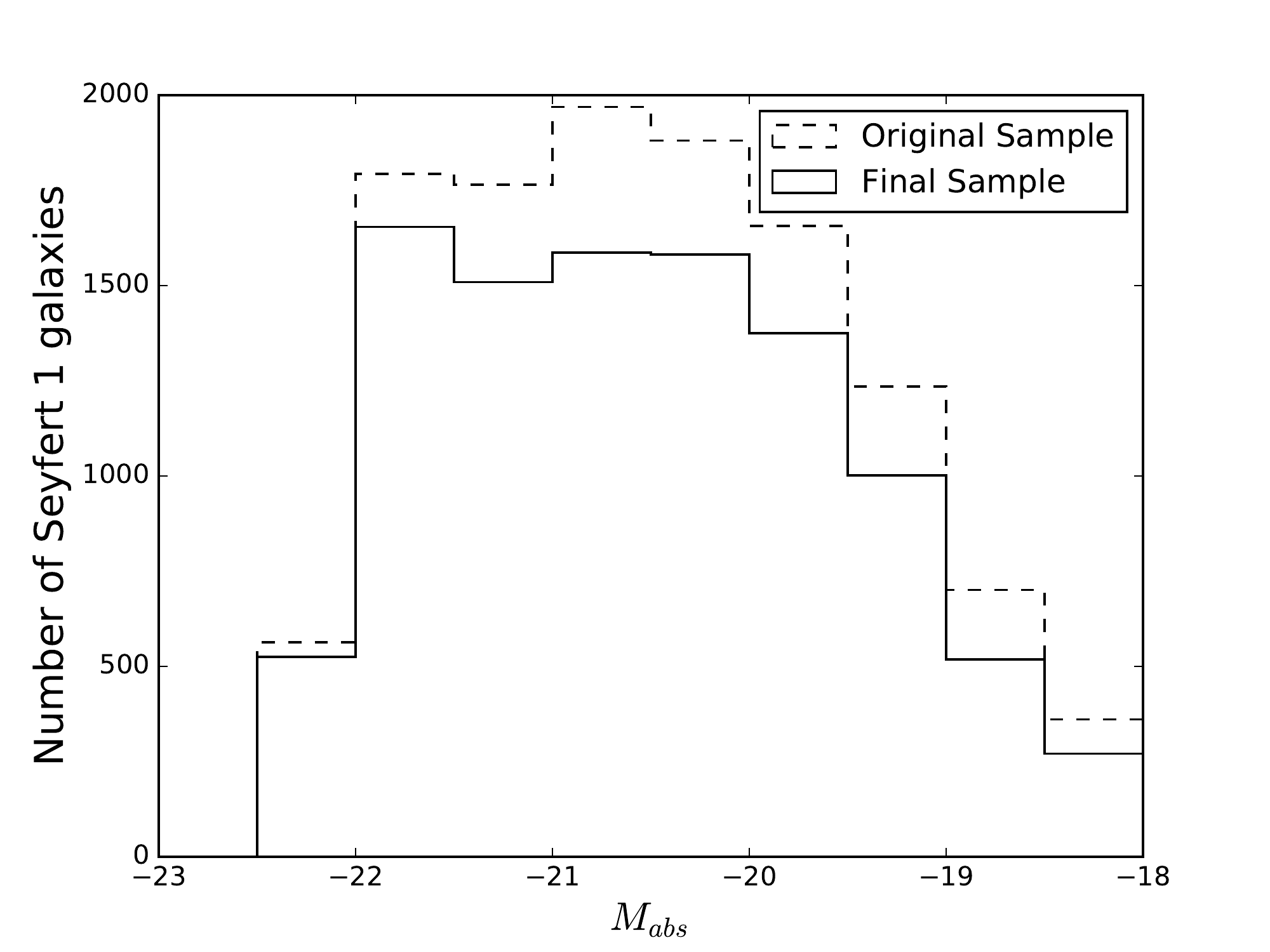}
		\caption{Histogram of number of galaxies per absolute magnitude bin. \citet{VeronCetty2010} impose a cut-off of -22.5 absolute magnitude for the distinction between QSOs and AGNs, which is why there are no objects with absolute magnitude less than -22.5}
		\label{Mabs hist}
	\end{figure}
	
	A vast majority of the Seyfert 1 sample lies within redshift $z \leq
	0.8$. Therefore, we have further truncated the sample to only include
	objects with redshift $z < 0.8$, as the number of objects beyond this
	range are too few in number, and their contribution to the bluest part
	of the rest-frame composite spectrum will be very noisy. After
	imposing all these conditions  on sample selection, we are left with a
	final sample of 10,112 Seyfert 1 galaxies.
	
	This final sample has a mean redshift $z_{mean} = 0.263 \pm
	0.149$, with a mean absolute magnitude $M_{B} = -20.543 \pm
	1.138$. In Figure~\ref{redshift hist} we plot the histogram of
	redshift for both the original sample of 12,129 SDSS objects and the
	final sample of 10,112 Seyfert 1 galaxies. Similarly, in
	Figure~\ref{Mabs hist} we plot the histogram of the absolute
	magnitude of the original and final sample. These values of redshift
	and absolute magnitude were obtained from the same Table\_AGN that
	the sample was drawn from. The absolute magnitude is k-corrected
	assuming an optical spectral index $\alpha$ (defined as $ S \propto
	\nu^{-\alpha}$) equal to 0.3. In order to test whether the selection
	cuts to the sample changes its redshift and absolute magnitude
	distribution, we performed a two-sample Kolmogorov-Smirnov (KS) test
	with the original and final sample distributions with respect to
	redshift and absolute magnitude. A small p-value (< 0.01) is required to statistically reject the null hypothesis which states that the two distributions of the final and original
	sample are drawn from the same underlying population. This test returned a p-value of
	0.8898 for the redshift distributions, and 0.1473 for the absolute magnitude samples. This indicates that the null
	hypothesis cannot be
	rejected with high significance considering either redshift or
	absolute magnitude. We may then assume that our final sample is not
	statistically biased with respect to the original sample. We now
	proceed to construct a median and geometric mean composite Seyfert 1
	spectrum from the final sample of 10,112 galaxies.

	\section{Generating the Composite Seyfert 1 Spectrum} \label{methodology}
	
	We wish to compare the composite Seyfert 1 spectrum with the composite
	quasar spectrum. The best way to do so would be to compare their
	respective spectral indices, and check for common emission features in
	the spectra. To do this, it is necessary to obtain a global continuum
	shape for our composite spectrum, while also preserving the emission
	lines in the spectrum.  Two different methods for combining spectra,
	viz. median spectrum and geometric mean spectrum are required to
	optimally obtain measurements of the continuum and of emission
	lines. The median stacking method preserves the relative fluxes of the
	emission features, while the geometric mean spectrum preserves the
	global shape of the continuum, and hence provides an accurate spectral
	index for the continuum. Depending on whether the emission line
	properties or the continuum shape are to be studied, one of these
	spectra will be more suitable than the other. To stack the spectra, we
	adopted the methodology of \citet{VandenBerk2001}. For completeness,
	we summarize the steps in the process here.
	
	There are three main steps involved in the generation of a composite
	spectrum, viz. shifting the spectrum to rest frame, scaling of
	spectrum, and stacking the shifted and scaled spectrum.
	
	At the time of \citet{VandenBerk2001}, the SDSS redshift pipeline was
	not sufficiently developed, and consequently, it was necessary to
	determine the redshift of each spectrum by examining the shift in the ubiquitous [OIII] line in the quasar spectra. However, since then, a
	robust redshift pipeline has become available with appropriate error
	flags (zWarning) indicating possible problems with measuring the
	redshift. Since we have already excluded galaxies with non-zero
	zWarning flags from our sample, we can directly use the redshift
	obtained from the SDSS database.
	
	Using the SDSS redshifts, all the spectra were
	shifted to rest frame.  Each spectrum was then re-binned into bins of
	width $1 \angstrom$, while conserving flux. The spectra were then
	ordered by redshift, and the first spectrum was arbitrarily
	scaled. The other spectra were scaled in order of redshift to the
	average of the flux density in the common wavelength region of the
	mean spectrum of all lower redshift spectra. The final spectrum was
	obtained by finding the median flux density in each bin of the
	shifted, re-binned and scaled spectra.
	
\begin{figure}
		\centering
		\includegraphics[width = \columnwidth]{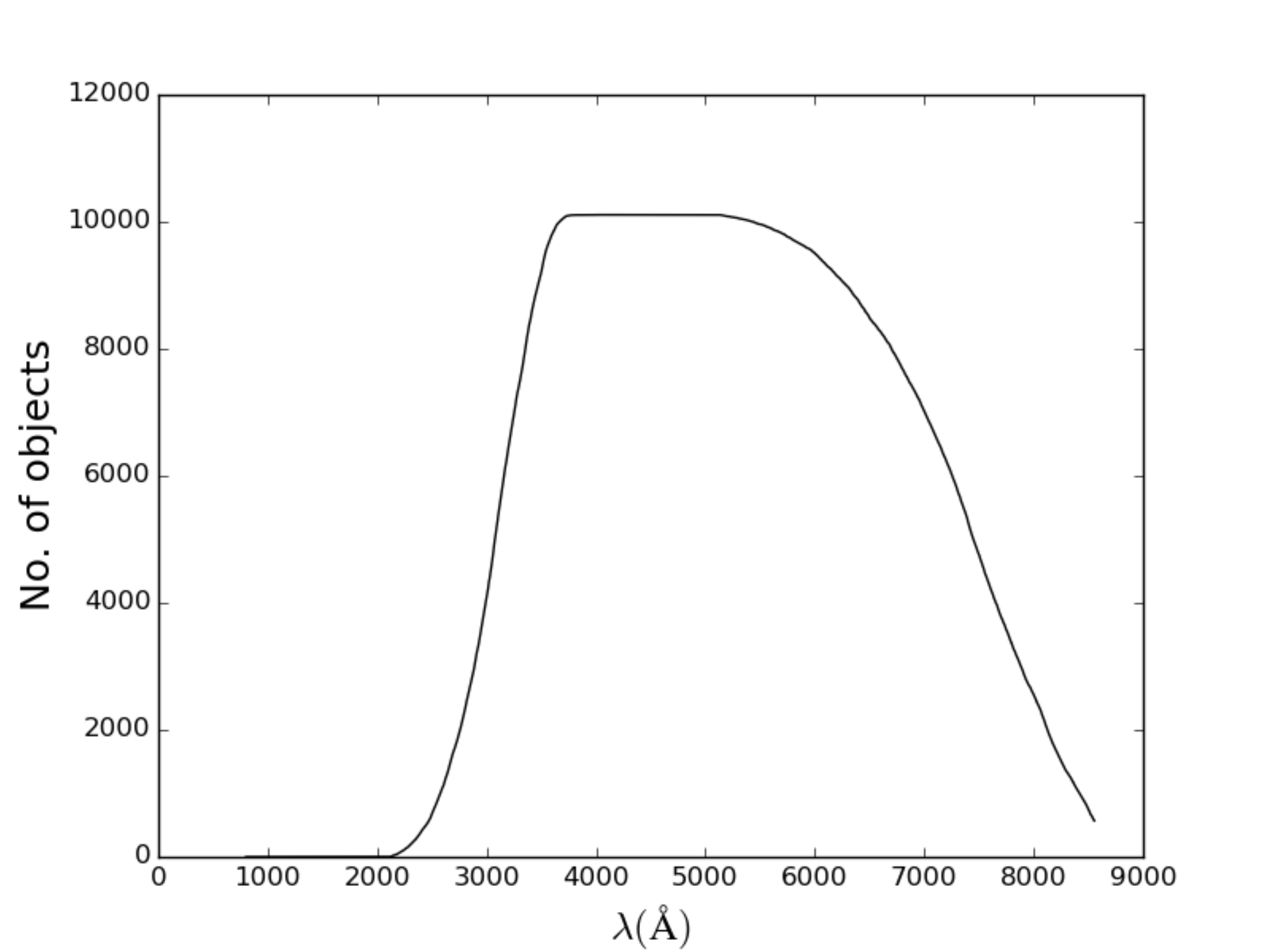}
		\caption{Number of spectra contributing to each $1\angstrom$ bin in the median spectrum.}
		\label{spectra per bin}
		
	\end{figure}

\begin{figure}
		\centering
		\includegraphics[width = \columnwidth]{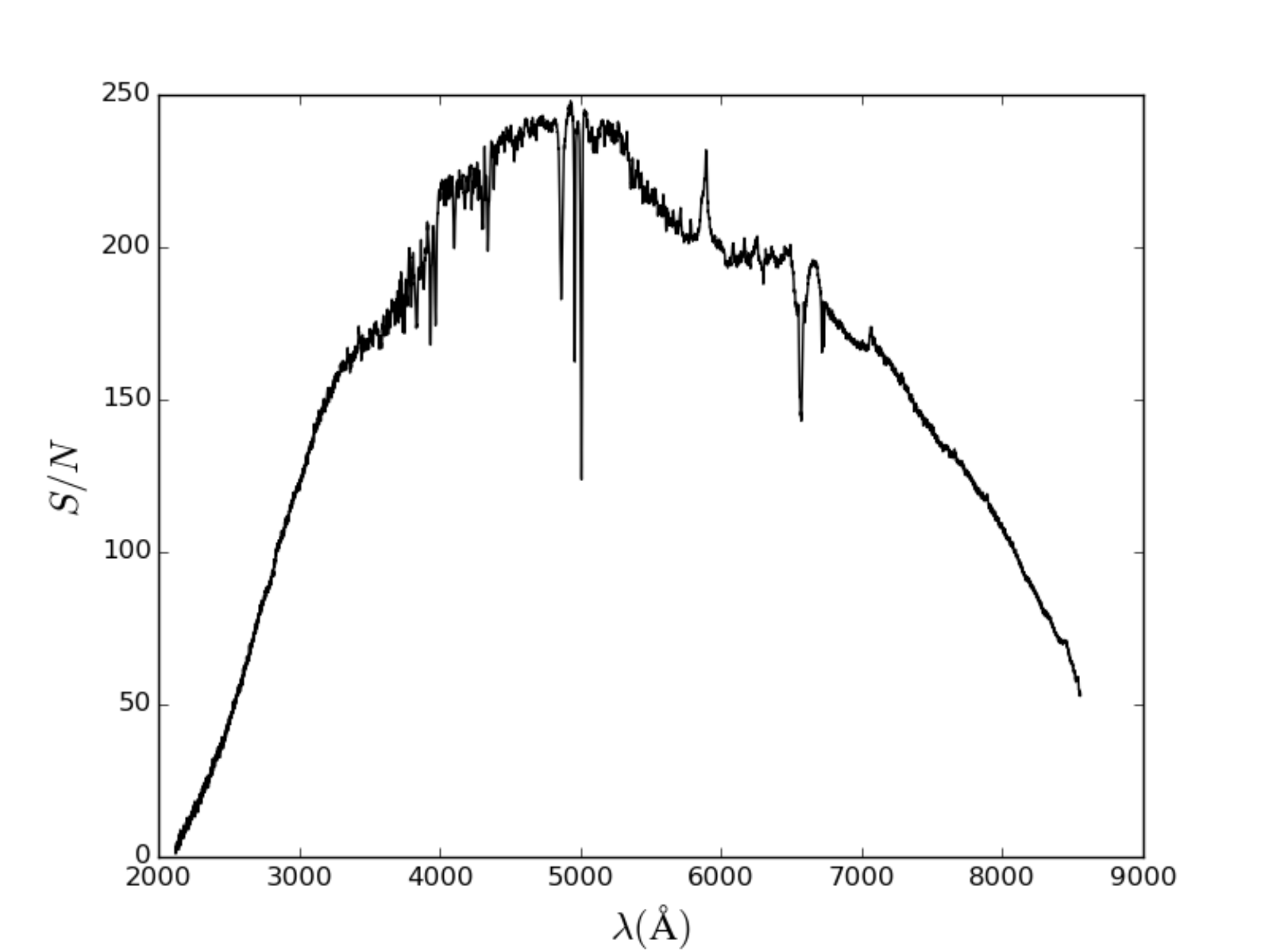}
		\caption{The signal-to-noise ratio for the median composite Seyfert 1 spectrum as a function of wavelength. The deep troughs correspond to emission lines in the spectrum. They are generated due to a small semi-interquartile range for the fluxes of emission lines due to approximately equal strengths of these emission lines in individual spectra.}
		\label{snr}
		
	\end{figure}

	The median spectrum thus obtained is shown in Figure~\ref{seyfert median
		spectrum}. The number of spectra that contribute to each bin are
	shown in Figure~\ref{spectra per bin}. An error array was also computed
	to determine the $1\sigma$ deviation values for the obtained
	spectrum. This was done by calculating the 68\% semi-interquartile
	range of the flux densities, divided by the square root of the number
	of spectra contributing to each bin. The corresponding signal-to-noise
	per bin obtained is shown in Figure~\ref{snr}.
	
	\begin{figure}
		\centering
		\includegraphics[width = \columnwidth]{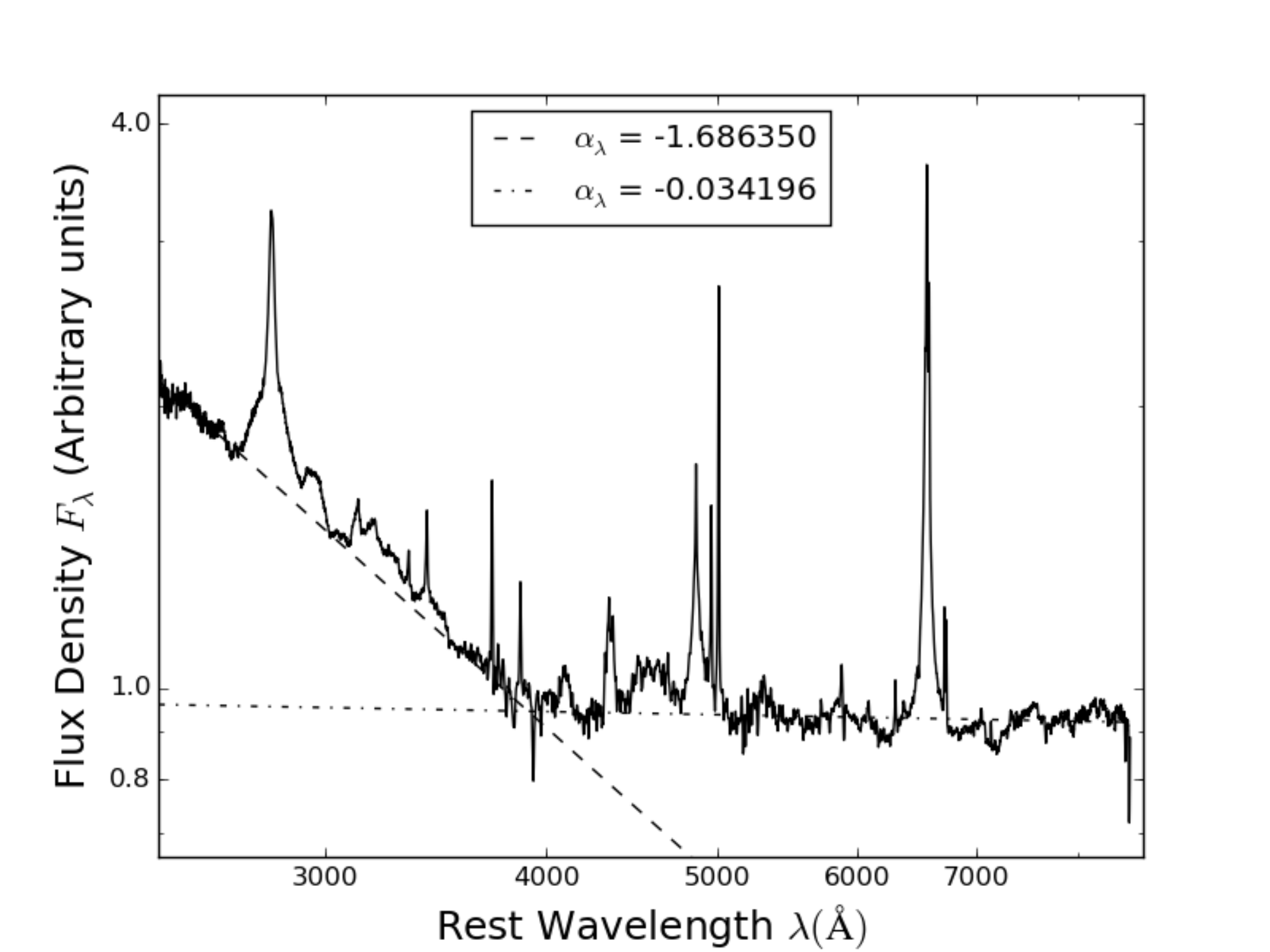}
		\caption{Seyfert 1 composite spectrum generated using median stacking. The figure has a log-log scale, and the power law fits to the blue and red region of the continuum are shown as dashed and dot-dashed lines respectively.}
		\label{seyfert median spectrum}
		
	\end{figure}
	
\begin{figure}
		\centering
		\includegraphics[width = \columnwidth]{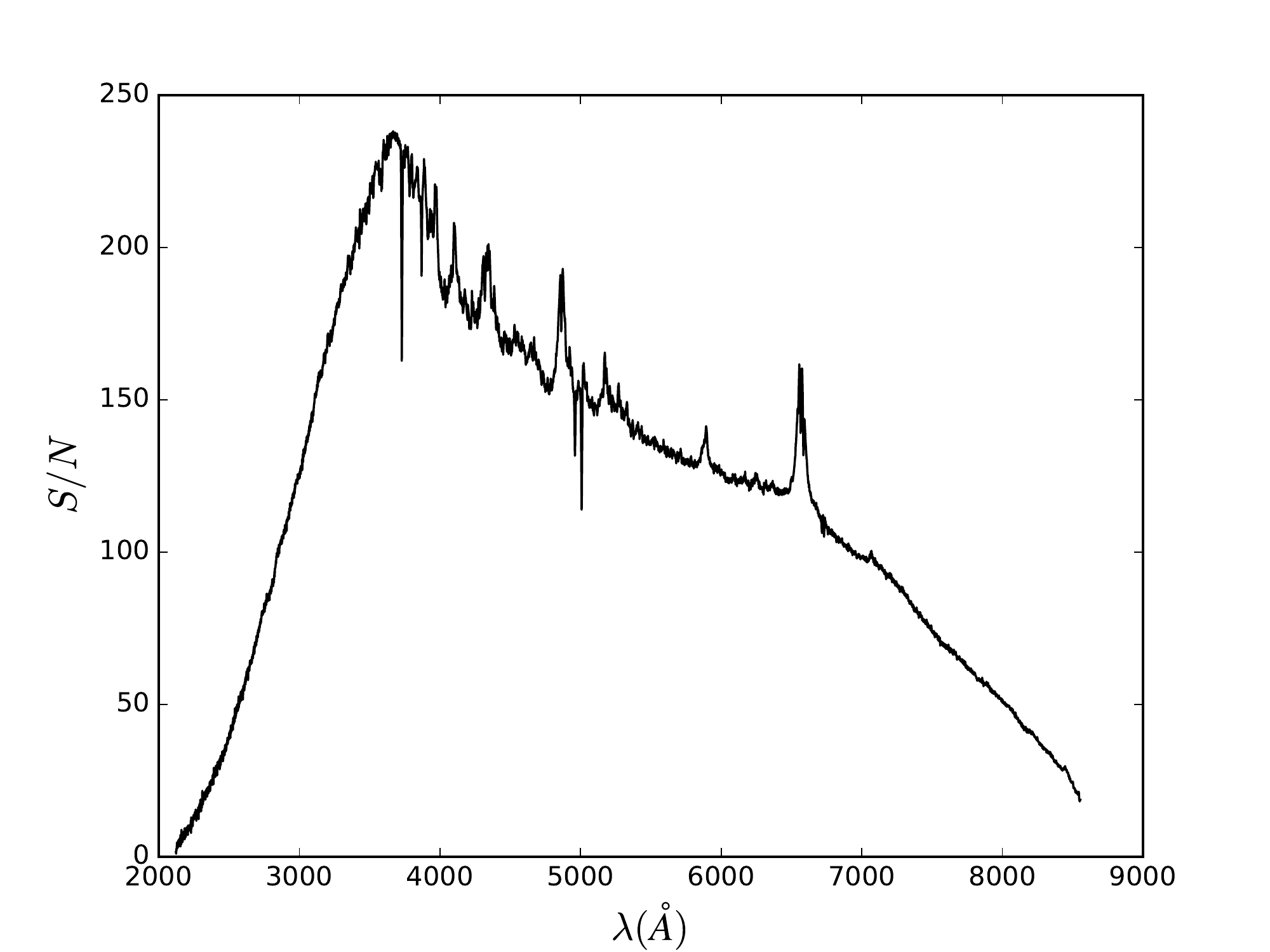}
		\caption{The signal-to-noise ratio for the geometric mean Seyfert 1 composite spectrum.}
		\label{GM_SNR}
	\end{figure}

	To produce the geometric mean composite spectrum, the shifted and
	re-binned spectra were normalized to unit average flux density
	over the rest-wavelength interval $3600-3700 \angstrom$. This interval
	was chosen because it contains no strong narrow emission lines. This
	constraint results in a total of 9812 objects contributing to the
	geometric mean spectrum. The number of spectra contributing to each
	bin in the geometric mean spectrum are shown in Figure~\ref{spectra
		per bin GM}. Additionally, a small number ($\lesssim 0.4$ \%)
	of pixels in the SDSS data may have negative flux values, which
	cannot be used while computing a geometric mean. Such pixels were
	masked and not used when the geometric mean was computed. This
	"clipped" geometric mean spectrum is shown in Figure~\ref{gm
		spectrum}. The corresponding error, computed in the same way as for
	the median spectrum, is shown in Figure~\ref{GM_SNR}.
	
	Both these composite spectra are available as an electronic table in the
	online version of this paper. The wavelength, flux, and error for a subset of the spectra are shown in Table~\ref{sample_table}.
	
	\begin{table*}
		\title{Median Seyfert 1 Composite Spectrum}
		\begin{tabular}{lll}
			\hline
			Wavelength ($\angstrom$) & Flux density (arbitrary units) & Error (arbitrary units) \\
			\hline
			2122.5 & 1.77712735783 & 0.923829810185 \\
			2123.5 & 0.84448103612 & 0.84230030616 \\
			2124.5 & 2.57595970771 & 0.856977684361 \\
			2125.5 & 5.21397283377 & 1.50567701417 \\
			2126.5 & 1.71139567619 & 0.687347713267 \\
			2127.5 & 1.95726066225 & 0.64081172443 \\
			2128.5 & 1.59422421416 & 0.651461404835 \\
			2129.5 & 1.49231370312 & 0.32758731566 \\
			2130.5 & 2.10427057004 & 0.730592977066 \\
		\end{tabular}
		\caption{The composite median spectrum in tabular form. Only a small number of rows are shown here for guidance about its structure. The entire table is available in electronic form in the online version of this paper.  The geometric mean spectrum, and the spectra binned in different redshift bins are also available online in the same format.}
		\label{sample_table}
	\end{table*}
	
	\begin{figure}
		\centering
		\includegraphics[width = \columnwidth]{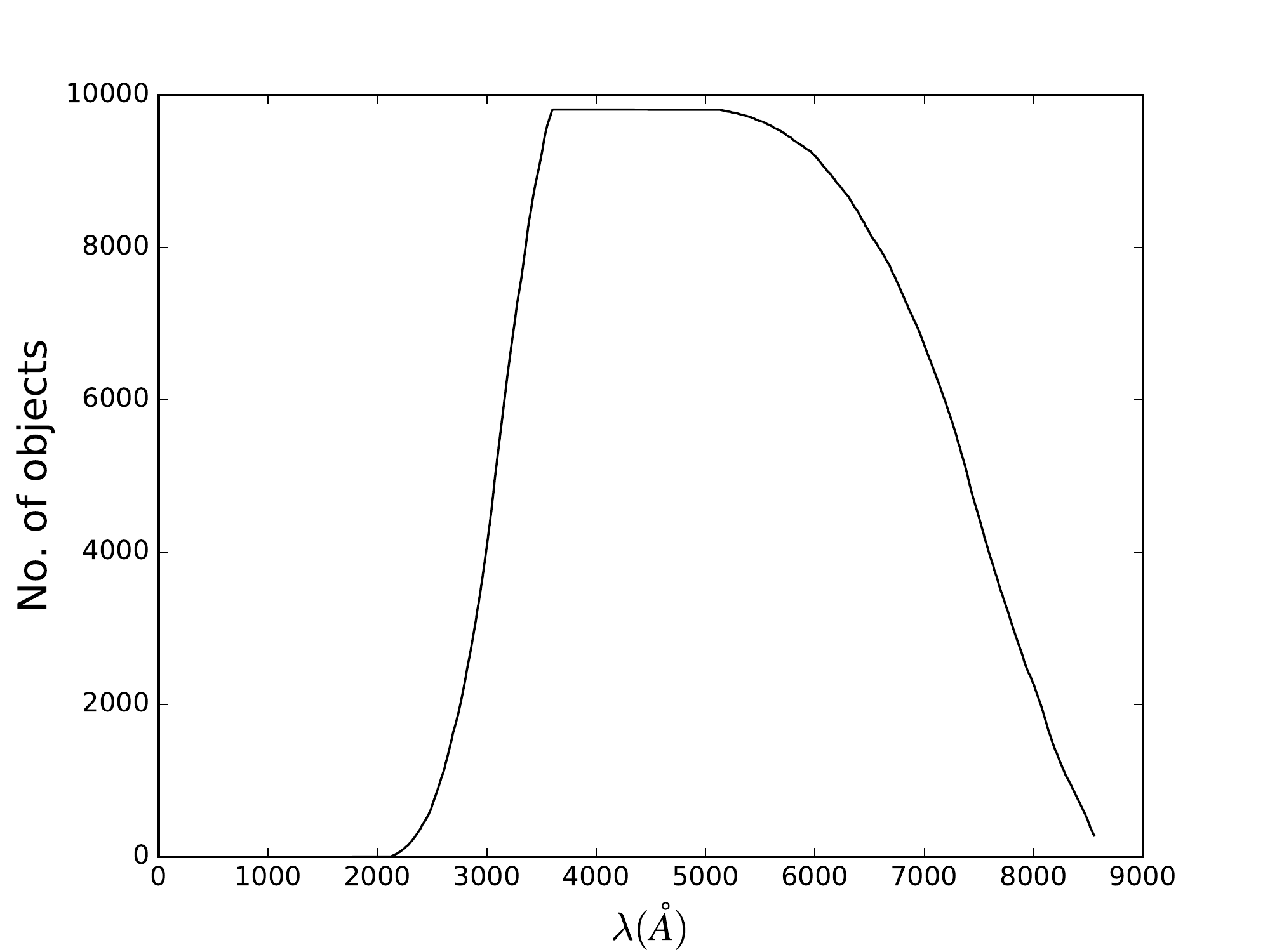}
		\caption{Number of spectra contributing to each $1 \angstrom$ bin in the geometric mean spectrum.}
		\label{spectra per bin GM}
	\end{figure}
	
	\begin{figure}
		\centering
		\includegraphics[width = \columnwidth]{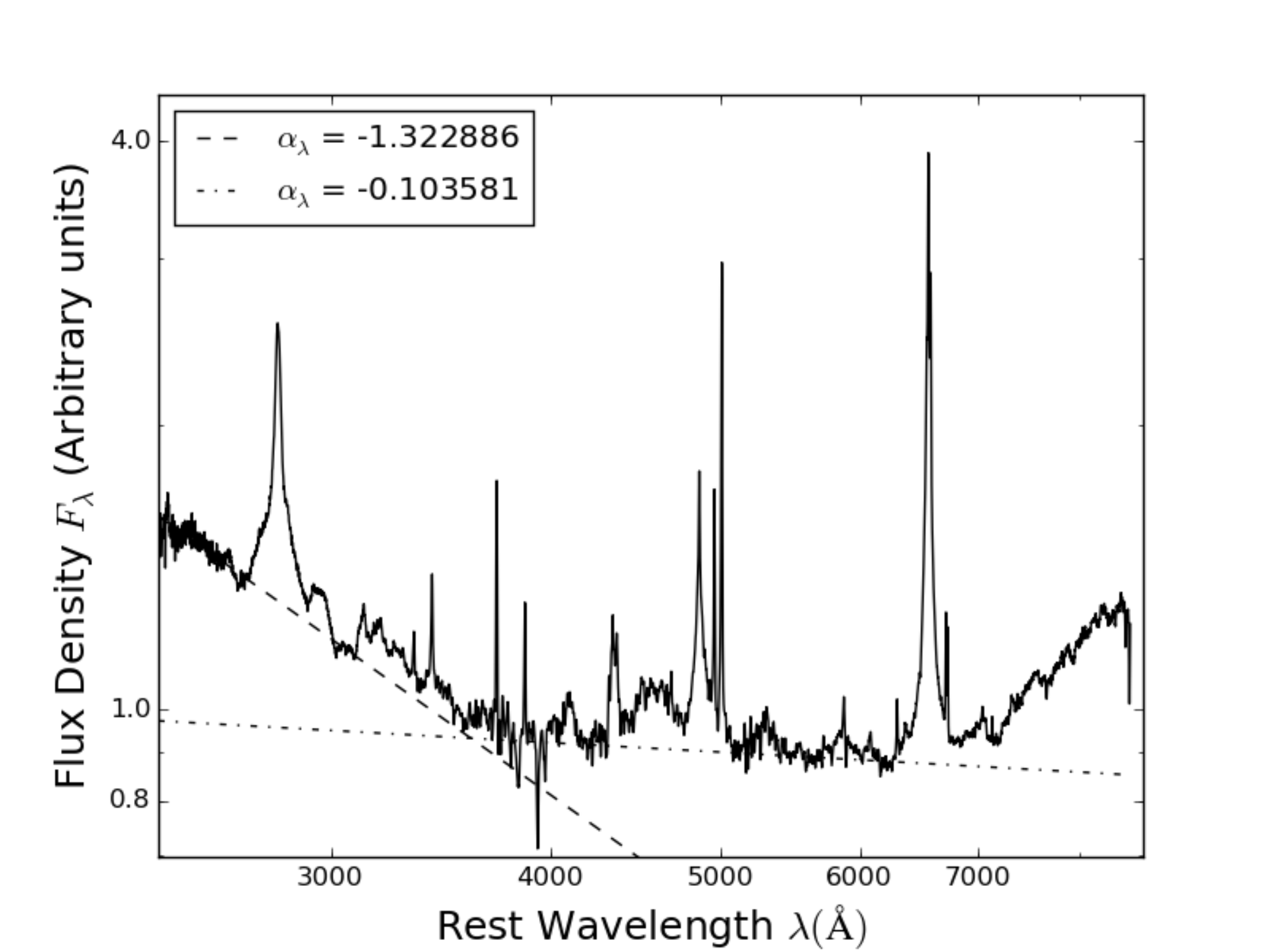}
		\caption{Seyfert 1 composite spectrum generated using geometric mean. The figure has log-log scale, and power law fits to the blue and red region of the continuum are shown as dashed and dot-dashed lines respectively.}
		\label{gm spectrum}
		
	\end{figure}
	
	\section{Continuum, Emission, and Absorption Features}
	\label{features}
	
	\subsection{The Continuum}
	\label{continuum}
	
	AGN continuum spectra are usually modeled as power-laws,
	
	\begin{equation}
	\displaystyle f_{\lambda} = \lambda ^ {\alpha}
	\end{equation}
	where $\alpha$ is the spectral index
	
	The geometric mean composite spectrum is shown in Fig. \ref{gm
		spectrum} on a log-log plot, along with the continuum power-law fit,
	which appears as a straight line. Because of contamination from
	several strong emission lines, it is difficult to fit a curve to the entire geometric mean spectrum by conventional techniques such as a least-square fit. Therefore,
	we choose two wavelength regions, situated as far away from each other as possible, which are not contaminated by strong emission lines. We then fit a power law to the data in these two regions using a non-linear least squares fit. The regions that we chose were $2200\angstrom - 2220\angstrom$ and $3020\angstrom - 3040\angstrom$. Using this method, we also measure the $1\sigma$ deviation from the median slope. The slope obtained for the geometric mean spectrum bluewards of $H\beta$
	line is $\alpha = -1.32$. The corresponding $1\sigma$ value is
	0.09.  At wavelengths longer than that of the $H\beta$ line, the flux
	rises above that predicted from the power-law fit. As a result, a
	different power-law fit is required for this region. The procedure for
	fitting a curve to this region is the same as described above, but with a region of $4105\angstrom - 4125\angstrom$ and $6200\angstrom - 6220\angstrom$ chosen for performing the fit. This
	yields a spectral index of $\alpha = -0.1$, with a $1\sigma$ deviation
	value of 0.138.
	
	The corresponding values obtained in \citet{VandenBerk2001} for the
	blue and red part of the geometric mean quasar spectrum are
	$\alpha = -1.56$ and $\alpha = 0.45$
	respectively. These values differ significantly from those that are
	obtained from the composite Seyfert 1 spectrum.
	
	We have also fit a continuum slope to the median composite
	spectrum. The corresponding values are $\alpha = -1.68$ for the blue
	portion of the spectrum, and $\alpha = -0.03$ for the red portion of
	the spectrum. These can be compared with the values of $\alpha =
	-1.54$ and $\alpha = -0.42$ from Vanden Berk median composite quasar spectrum respectively.
	
	\subsection{Emission Features}
	\label{emission}
	
\begin{figure*}
		\includegraphics[width = \textwidth]{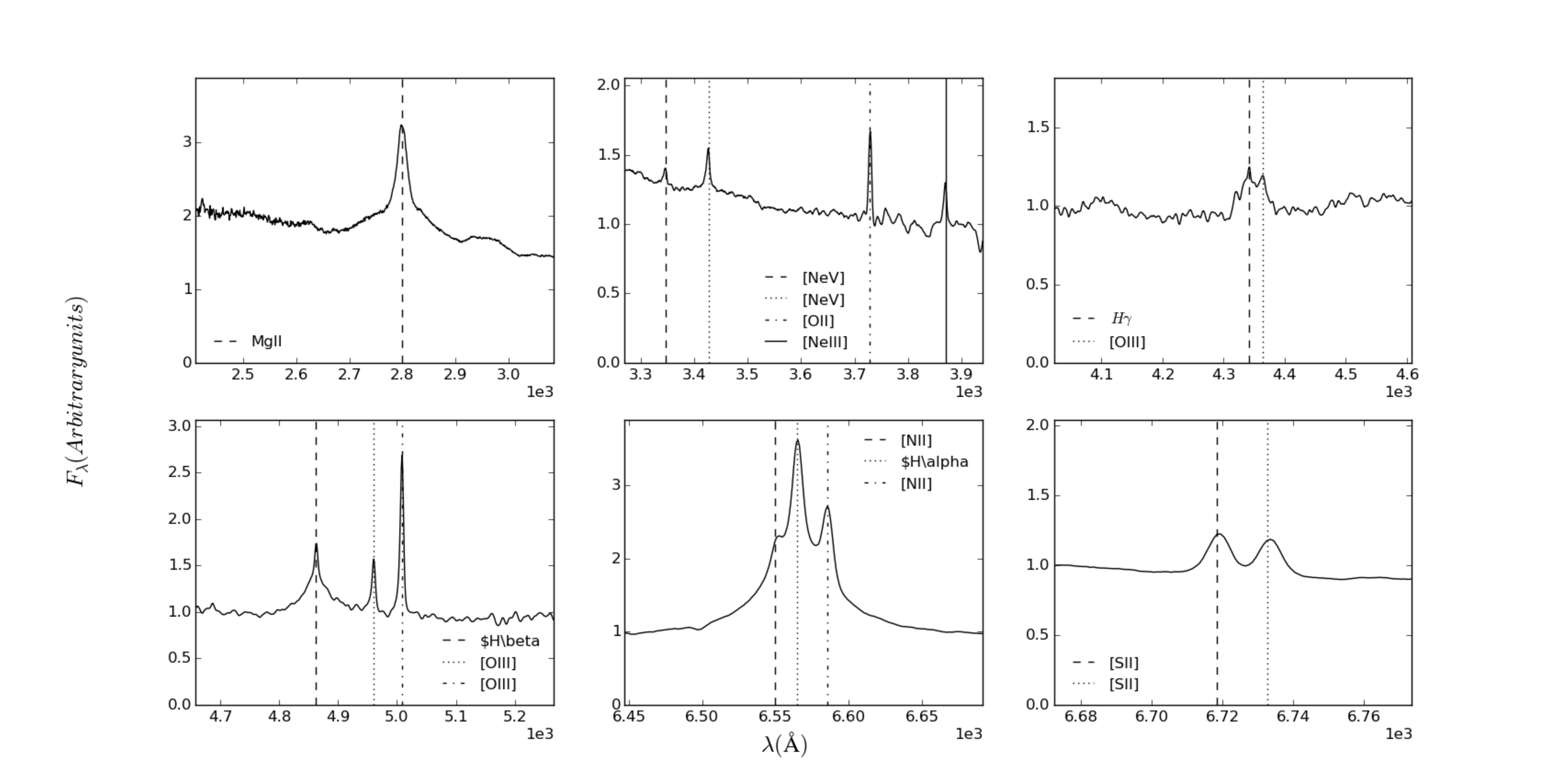}
		\caption{Strong emission line features in the composite Seyfert 1 spectrum}
		\label{emission_lines}
	\end{figure*}

	Due to the high SNR for our median composite spectrum, we are able to clearly
	detect many emission lines in the median Seyfert 1
	spectrum. Fig. \ref{emission_lines} shows some of the most clearly
	visible lines in the median spectrum. The characteristic broad lines of
	$MgII$ and $H\alpha$ are clearly visible, along with the narrower
	lines of $[OIII], [OII], H\beta$ etc. In addition, $FeII \ \textrm{and} \ FeIII$ complexes can also be seen in the median spectrum. Table \ref{emission_table} lists the center wavelengths,
	measured equivalent widths, and fluxes of these lines. To compute
	these properties, we made use of the software package Specview. We list in Table~\ref{emission_table} all the emission lines clearly detectable above the continuum in the median composite spectrum. We list a total of 39 emission lines, some of which are blended together in the spectrum. In the cases where we could not separate the lines, we report the flux and equivalent width of the blended line. In comparison, the median composite quasar spectrum contains 53 emission lines in the wavelength region greater than $2800 \angstrom$. We did not detect any line that is not already present in the composite quasar spectrum of \citet{VandenBerk2001}.
	
	\begin{table*}
		\begin{tabular}{lllll}
			\hline
			$\lambda_{obs}$ & Amplitude & Eq. Width & ID  & $\lambda_{lab}$ or Multiplet \\
			(\angstrom) & (Arbitrary units) & (\angstrom) &  & (\angstrom) \\
			\hline
			2797.48 $\pm$ 3.12 & 64.272 $\pm$ 0.784 & 36.10 $\pm$ 1.16 & MgII & 2798.75 \\
			3129.81 $\pm$ 2.69 & 2.273 $\pm$ 0.090 & 1.55 $\pm$ 0.06 & OIII & 3133.70 \\
			& & & FeII & Opt82 \\ 
			3344.74 $\pm$ 1.23 & 0.954 $\pm$ 0.055 & 0.74 $\pm$ 0.04 & [NeV] & 3346.82 \\
			3425.69 $\pm$ 0.66 & 3.358 $\pm$ 0.052 & 2.73 $\pm$ 0.04 & [NeV] & 3426.84 \\
			3728.62 $\pm$ 1.27 & 4.169 $\pm$ 0.220 & 4.02 $\pm$ 0.29 & [OII] & 3728.48 \\
			3760.55 $\pm$ 0.41 & 1.926 $\pm$ 0.027 & 2.00 $\pm$ 0.03 & [FeVII] & 3759.99 \\
			3783.77 $\pm$ 0.83 & 1.930 $\pm$ 0.045 & 2.03 $\pm$ 0.05 & FeII & Opt15 \\
			3815.55 $\pm$ 0.97 & 1.767 $\pm$ 0.037 & 1.91 $\pm$ 0.04 & FeII & Opt14 \\
			3869.12 $\pm$ 1.35 & 2.020 $\pm$ 0.113 & 2.00 $\pm$ 0.12 & [NeIII] & 3869.85 \\
			3891.14 $\pm$ 1.21 & 0.140 $\pm$ 0.017 & 0.14 $\pm$ 0.02 & HeI & 3889.74 \\
			& & & H8 & 3890.15 \\
			4072.19 $\pm$ 3.87 & 0.389 $\pm$ 0.082 & 0.40 $\pm$ 0.08 & [FeV] & 4072.39 \\
			4102.70 $\pm$ 0.88 & 0.206 $\pm$ 0.020 & 0.20 $\pm$ 0.02 & H$\delta$ & 4102.89 \\
			4138.05 $\pm$ 2.32 & 0.305 $\pm$ 0.029 & 0.32 $\pm$ 0.09 & FeII & Opt27 \\
			& & & FeII & Opt28 \\
			4320.82 $\pm$ 1.56 & 1.277 $\pm$ 0.088 & 1.34 $\pm$ 0.10 & [FeII] & Opt21F \\
			& & & FeII & Opt32 \\
			4340.93 $\pm$ 0.87 & 5.313 $\pm$ 0.121 & 5.55 $\pm$ 0.18 & H$\gamma$ & 4341.68 \\
			4364.21 $\pm$ 0.84 & 3.482 $\pm$ 0.098 & 3.61 $\pm$ 0.13 & [OIII] & 4364.44 \\
			4687.15 $\pm$ 2.28 & 0.681 $\pm$ 0.085 & 0.68 $\pm$ 0.36 & HeII & 4687.02 \\
			4861.79 $\pm$ 0.78 & 21.859 $\pm$ 0.179 & 22.71 $\pm$ 0.53 & H$\beta$ & 4862.68 \\
			4931.32 $\pm$ 1.31 & 1.649 $\pm$ 0.095 & 1.74 $\pm$ 0.11 & FeII & Opt42 \\
			4958.15 $\pm$ 0.90 & 7.169 $\pm$ 0.135 & 7.59 $\pm$ 0.23 & [OIII] & 4960.30 \\
			5007.77 $\pm$ 0.39 & 15.330 $\pm$ 0.135 & 16.39 $\pm$ 0.41 & [OIII] & 5008.24 \\
			5159.39 $\pm$ 0.92 & 0.489 $\pm$ 0.021 & 0.54 $\pm$ 0.02 & [FeVII] & 5160.33 \\
			5179.48 $\pm$ 0.69 & 0.454 $\pm$ 0.020 & 0.53 $\pm$ 0.02 & [FeVI] & 5177.48 \\
			5201.55 $\pm$ 0.78 & 0.589 $\pm$ 0.023 & 0.64 $\pm$ 0.03 & [NI] & 5200.53 \\
			5725.23 $\pm$ 1.49 & 0.590 $\pm$ 0.027 & 0.64 $\pm$ 0.03 & [FeVII] & 5722.30 \\
			5873.90 $\pm$ 1.00 & 1.723 $\pm$ 0.035 & 1.84 $\pm$ 0.04 & HeI & 5877.29 \\
			6083.20 $\pm$ 1.71 & 0.789 $\pm$ 0.034 & 0.86 $\pm$ 0.04 & [FeVII] & 6087.98 \\
			6302.72 $\pm$ 0.92 & 0.983 $\pm$ 0.043 & 1.08 $\pm$ 0.05 & [OI] & 6302.05 \\
			6372.78 $\pm$ 1.59 & 0.647 $\pm$ 0.033 & 0.69 $\pm$ 0.03 & [OI] & 6365.54 \\
			& & & [FeX] & 6376.30 \\
			6565.39 $\pm$ 0.09 & 43.063 $\pm$ 0.124 & 47.79 $\pm$ 0.95 & H$\alpha$ & 6564.61 \\
			6585.77 $\pm$ 0.08 & 22.575 $\pm$ 0.075 & 25.10 $\pm$ 0.38 & [NII] & 6585.28 \\
			6717.99 $\pm$ 0.51 & 3.205 $\pm$ 0.055 & 3.56 $\pm$ 0.07 & [SII] & 6718.29 \\
			6733.75 $\pm$ 0.93 & 2.592 $\pm$ 0.068 & 2.88 $\pm$ 0.09 & [SII] & 6732.67 \\
			7138.67 $\pm$ 1.30 & 0.338 $\pm$ 0.022 & 0.39 $\pm$ 0.02 & [ArIII] & 7137.80 \\
		\end{tabular}
		\caption{Emission lines detected in the composite Seyfert 1 spectrum. Multiplet designations for Fe are from \citet{VandenBerk2001}}
		\label{emission_table}
	\end{table*}
	
	\subsection{Absorption Features}
	\label{absorption}
	
\begin{figure*}
		\includegraphics[width = \textwidth]{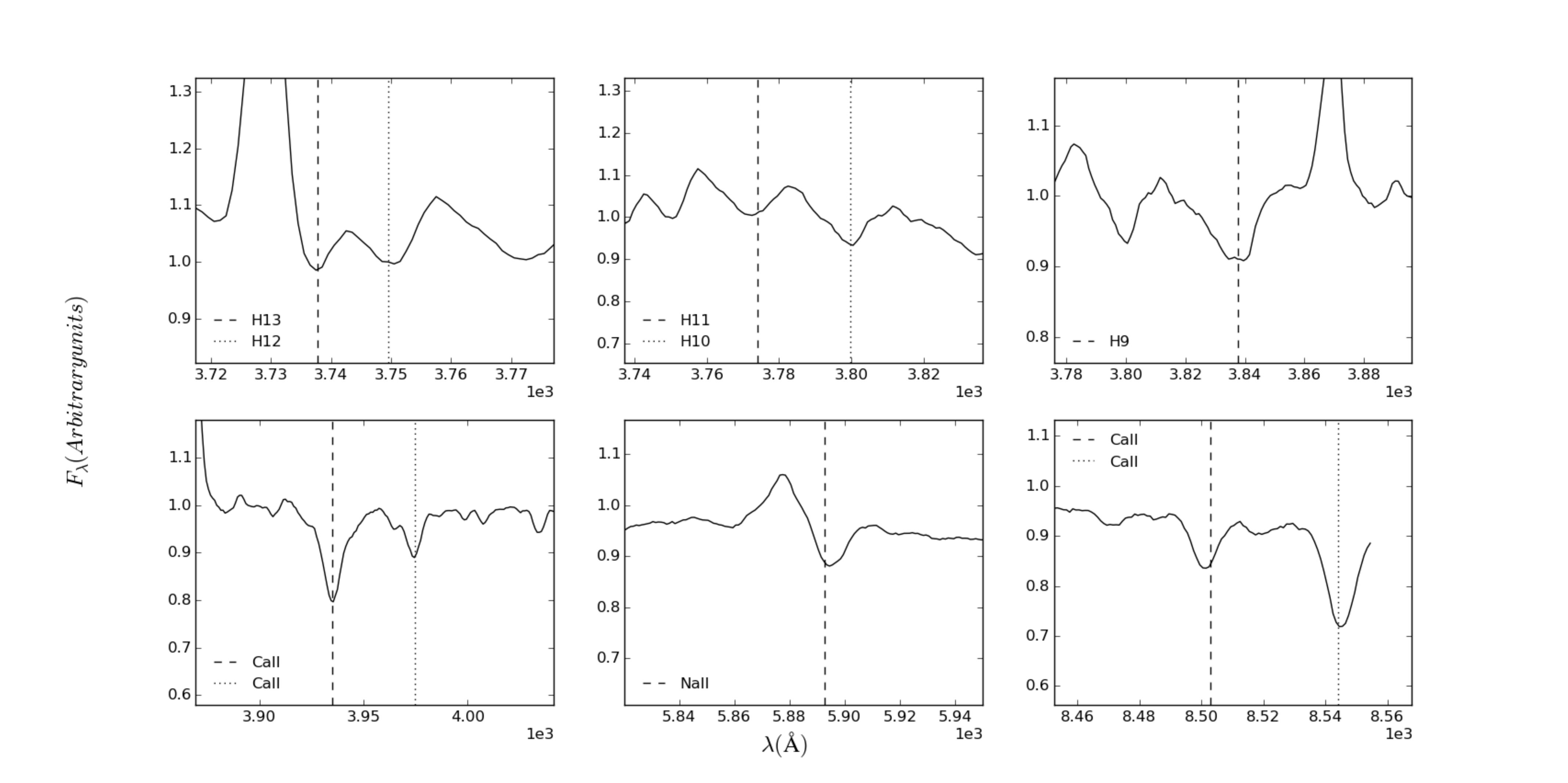}
		\caption{Absorption features in the composite Seyfert 1 spectrum. Note that the detection of H13 is uncertain due to its close proximity to the [OII] emission line.}
		\label{abs_lines}
	\end{figure*}

	We can also detect some absorption features in our median spectrum. The most
	prominent ones are displayed in Fig. \ref{abs_lines}. It is evident
	that the median composite Seyfert spectrum shows stellar absorption lines,
	particularly the Balmer lines and the Ca II $\lambda3968$ line. The
	presence of these lines is suggestive of stellar light contamination
	in the input Seyfert spectra. Just as for emission lines, we list the flux and equivalent widths of absorption lines clearly visible with respect to the continuum in Table \ref{absorption table}.
	
	\begin{table*}
		\begin{tabular}{lllll}
			\hline
			$\lambda_{obs}$ & Amplitude & Eq. Width & ID & $\lambda_{lab}$ \\
			(\angstrom) & (Arbitrary units) & (\angstrom) &  & (\angstrom) \\
			\hline
			3737.73 $\pm$ 3.18 & 0.507 $\pm$ 0.163 & 0.47 $\pm$ 0.15 & H13: & 3735.43 \\
			3751.09 $\pm$ 0.42 & 0.487 $\pm$ 0.021 & 0.45 $\pm$ 0.02 & H12 & 3751.22 \\
			3771.27 $\pm$ 2.41 & 1.010 $\pm$ 0.069 & 0.93 $\pm$ 0.06 & H11 & 3771.70 \\
			3797.96 $\pm$ 2.52 & 1.281 $\pm$ 0.081 & 1.23 $\pm$ 0.08 & H10 & 3798.98 \\
			3835.03 $\pm$ 1.47 & 1.938 $\pm$ 0.059 & 1.91 $\pm$ 0.06 & H9 & 3836.47 \\
			3934.67 $\pm$ 0.77 & 2.792 $\pm$ 0.048 & 2.80 $\pm$ 0.05 & CaII & 3934.78 \\
			3974.28 $\pm$ 0.50 & 0.779 $\pm$ 0.021 & 0.78 $\pm$ 0.02 & CaII & 3969.59 \\
			5895.45 $\pm$ 0.63 & 0.759 $\pm$ 0.019 & 0.79 $\pm$ 0.02 & NaII & 5891.58 \\
			8501.27 $\pm$ 0.63 & 0.821 $\pm$ 0.072 & 0.88 $\pm$ 0.08 & CaII & 8500.36 \\
			8544.80 $\pm$ 1.38 & 1.655 $\pm$ 0.085 & 1.86 $\pm$ 0.10 & CaII & 8544.44 \\
		\end{tabular}
		\caption{Strong absorption lines detected in the composite Seyfert 1 spectrum. The colon (:) in front of H13 denotes an uncertain identification due to its proximity to the [OII] emission line.}
		\label{absorption table}
	\end{table*}
	
	\section{Comparison with Composite Quasar Spectrum}
	\label{compare with quasar}
	
\begin{figure}
		\includegraphics[width = \columnwidth]{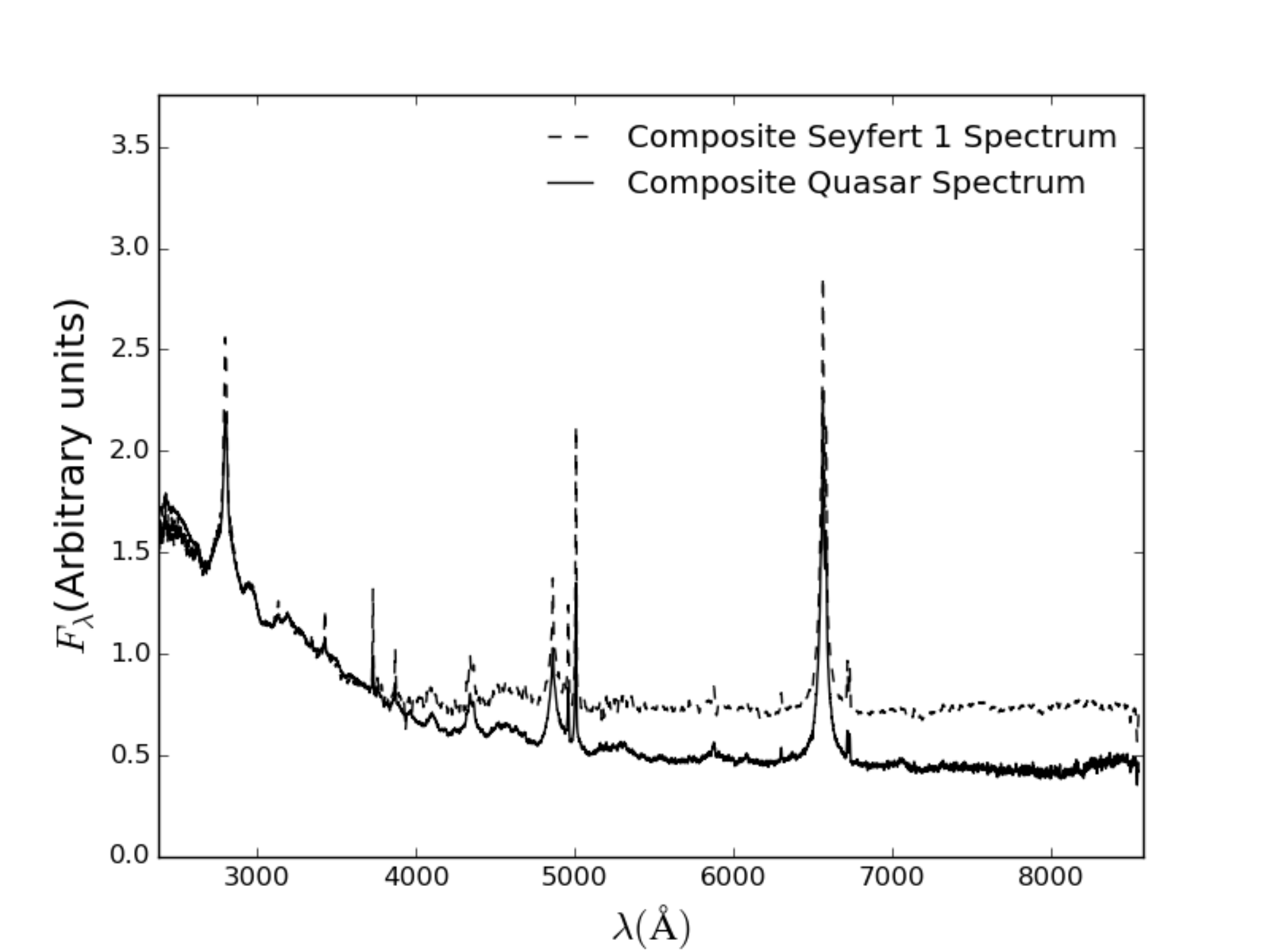}
		\caption{Comparison of composite Seyfert 1 spectrum with \citet{VandenBerk2001} composite quasar spectrum. Both spectra have been normalized to unit flux in the range $3020\angstrom-3100\angstrom$. These two spectra agree remarkably well in the blue portion of the spectrum, but are offset in the red portion of the spectrum.}
		\label{vdb vs s1}
	\end{figure}

\begin{figure}
		\includegraphics[width = \columnwidth]{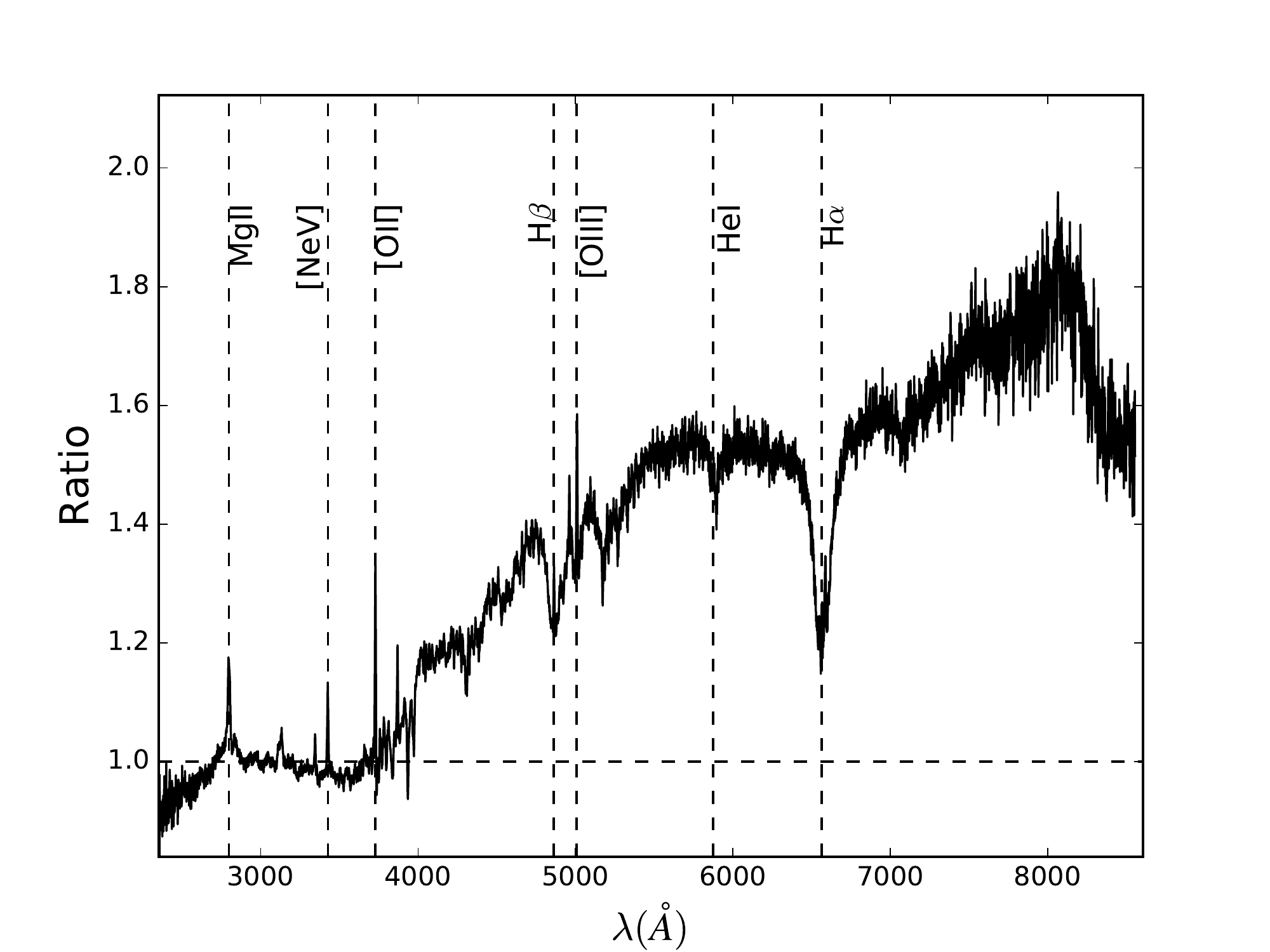}
		\caption{Ratio of median composite Seyfert 1 spectrum with median quasar composite spectrum. A few of the emission line features are labeled.}
		\label{ratio}
	\end{figure}

	Fig. \ref{vdb vs s1} compares our median Seyfert 1 composite spectrum with
	the median composite quasar spectrum from \citet{VandenBerk2001}. In producing
	this graph, both the spectra were normalized to unit flux in the range
	$3020\angstrom-3100\angstrom$ which does not contain any strong
	emission features. It is obvious that the spectral index for Seyfert
	1s is very similar to that of their quasar counterparts, both for the
	blue and red portion of the continuum. In order to see this better, Figure~\ref{ratio} shows the ratio of the median composite Seyfert 1 spectrum with the \citet{VandenBerk2001} median composite quasar spectrum. In order to compute this ratio, both composite spectra were normalized to unit flux in the range $3020\angstrom-3100\angstrom$, after which the median composite Seyfert 1 spectrum was divided by the median composite quasar spectrum. The resulting spectrum was again normalized to have unit flux in the same wavelength region. As is evident from Figure~\ref{ratio}, the continuum flux in the blue portion of the two composite spectra is approximately flat, while there is an offset in flux in the red portion of the spectrum. In addition to this, Figure~\ref{ratio} also shows the relative strength of emission lines between the two composite spectra. For instance, the composite Seyfert 1 spectrum has, on average, stronger MgII, [NeV] and [OIII] lines, while the composite quasar spectrum has stronger $H\beta$ and $H\alpha$ lines. We provide this ratio spectrum in electronic form in the online version of this paper.
	
	As can be seen, the main difference between these two spectra arises from the upturn
	in flux in the Seyfert 1 spectrum at approximately $4000\angstrom$. The reason for this upturn seems to be an increasing contribution of flux
	from the host galaxy. Our sample covers redshifts from almost 0 to
	0.8. Based on the angular diameter - redshift relation for our chosen cosmology, in the lowest redshift bin (average $z = 0.05$), the SDSS
	spectrograph fiber diameter of 3'' is small enough to cover only the
	nuclear region of the Seyfert galaxy, while at higher redshifts, it is
	big enough to cover almost the whole galaxy. However, the AGN in high
	redshift Seyferts are, on average, more luminous than their low
	redshift counterparts \citep{quasar_redshift_rel}. As a result, the
	light from high redshift Seyferts is dominated by the core AGN
	component, while that from the low redshift Seyferts shows significant
	contamination by host galaxy light.
	
\begin{figure*}
		\includegraphics[width = \textwidth]{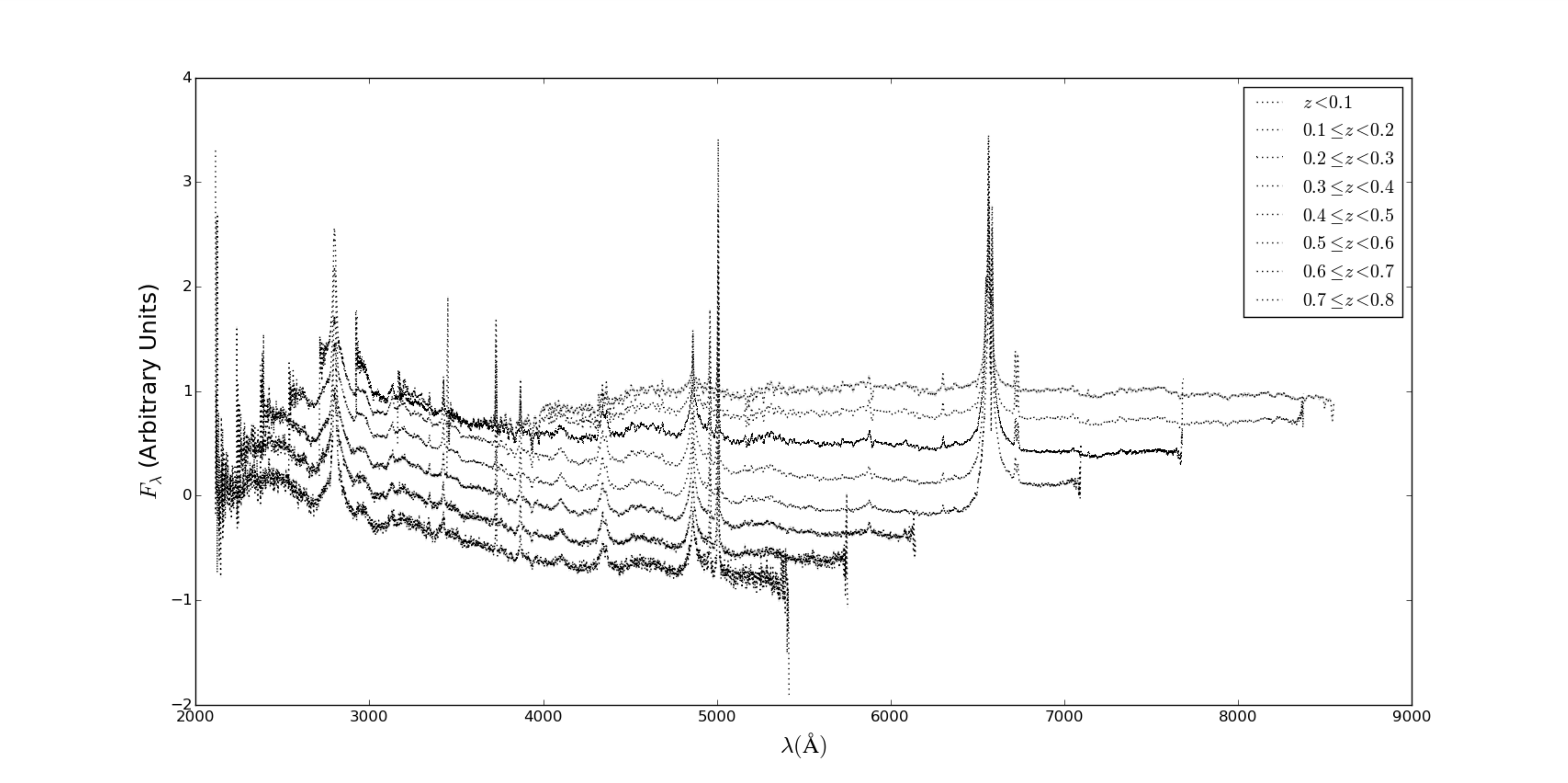}
		\caption{Variation in the composite Seyfert 1 spectrum with redshift. The different spectra are generated by median stacking the spectra into redshift bins of width 0.1. An offset in flux has been introduced in the stacked spectrum of successive bins for clarity. The lowest redshift bin is at the top, while higher redshift bins are progressively lower. Each spectrum is available in electronic form in the online version of this paper.}
		\label{complete redshift evolution}
	\end{figure*}

\begin{figure*}
		\centering
		\includegraphics[width = \textwidth]{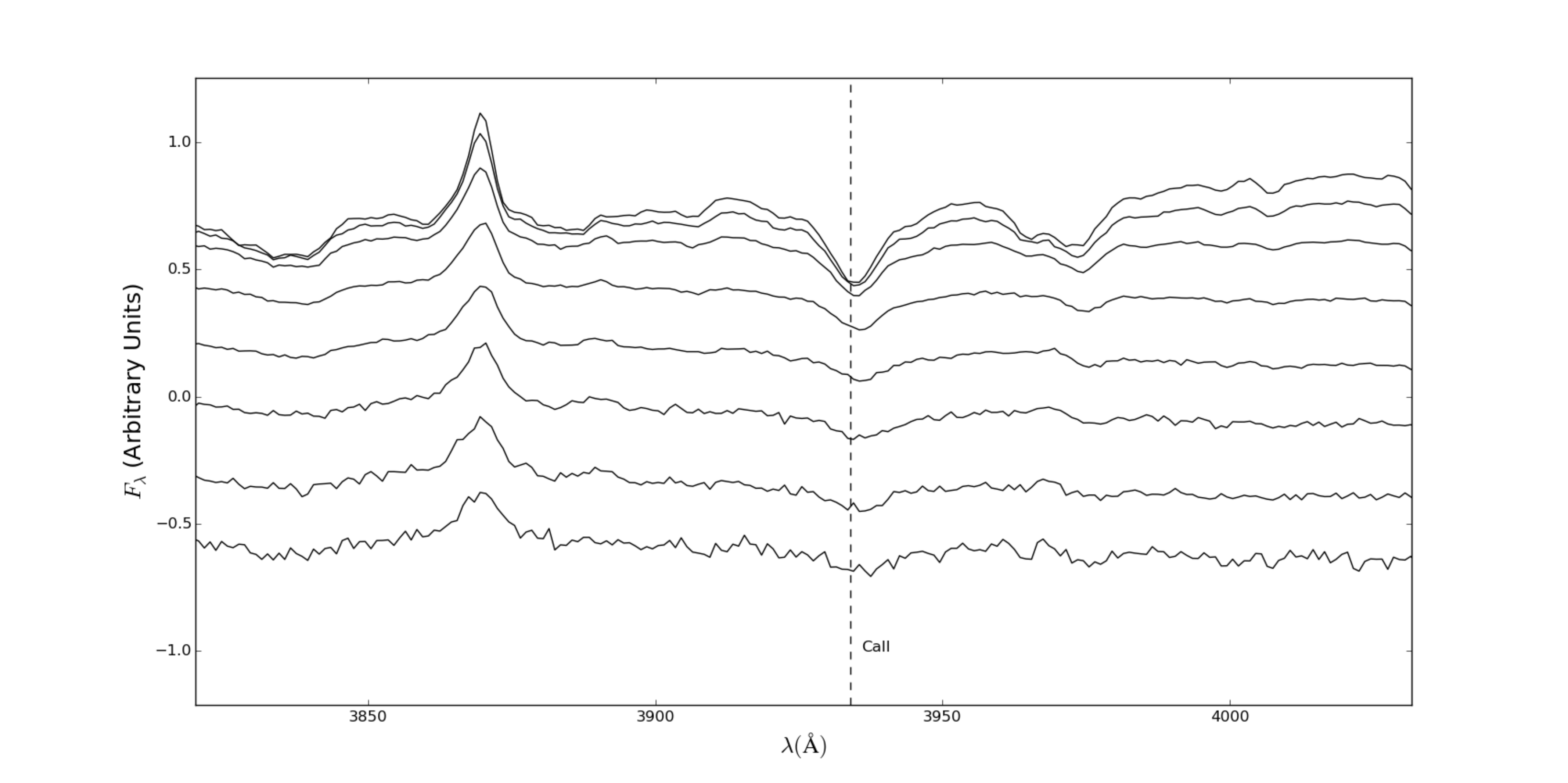}
		\caption{Variation in amplitude of the CaII absorption line with redshift. The spectra are ordered in increasing redshift from top to bottom. The topmost spectrum represents the bin $0 < z < 0.1$, the next one $0.1 < z < 0.2$, and so on till the last spectrum on the bottom, which represents $0.7 < z < 0.8$. Offsets have been introduced between each spectrum increase the visibility of each individual spectrum. The decrease in the amplitude is clearly visible with increase in redshift, with absorption lines very prominent in the lowest redshift bin, while  being essentially absent in the highest redshift bin.}
		\label{all_Ca}
	\end{figure*}

	To test whether this explanation is valid, we separately stacked
	spectra in redshift bins of width 0.1, in increasing order of
	redshift. A median composite spectrum was created for each of the bins
	in the same fashion as the complete median spectrum. The spectra
	produced are shown in Figure~\ref{complete redshift evolution}. If the
	above hypothesis is correct, we should see a decrease in the amplitude
	of absorption lines as we move from lower to higher redshift
	bins. Indeed, this is what we see when we look at the prominent
	absorption lines in the spectrum. This is illustrated for the CaII
	absorption line in Figure~\ref{all_Ca}, and the measured flux and equivalent widths are given in Table~\ref{CaII evolution}. This trend is visible for all
	the other absorption lines as well.
	
	\begin{table*}
		\begin{tabular}{lll}
			\hline
			Redshift bin & Amplitude & Eq. Width \\
			& (Arbitrary Units) & (\angstrom) \\
			\hline
			0.0 < z $\leq$ 0.1 & 4.337 $\pm$ 0.097 & 5.68 $\pm$ 0.15 \\
			0.1 < z $\leq$ 0.2 & 3.439 $\pm$ 0.133 & 3.83 $\pm$ 0.18 \\
			0.2 < z $\leq$ 0.3 & 2.833 $\pm$ 0.082 & 2.81 $\pm$ 0.09 \\
			0.3 < z $\leq$ 0.4 & 2.085 $\pm$ 0.072 & 2.06 $\pm$ 0.07 \\
			0.4 < z $\leq$ 0.5 & 1.796 $\pm$ 0.094 & 1.84 $\pm$ 0.10 \\
			0.5 < z $\leq$ 0.6 & 1.364 $\pm$ 0.099 & 1.45 $\pm$ 0.11 \\
			0.6 < z $\leq$ 0.7 & 1.046 $\pm$ 0.119 & 1.23 $\pm$ 0.14 \\ 
			0.7 < z $\leq$ 0.8 & 1.009 $\pm$ 0.156 & 1.24 $\pm$ 0.19 \\
		\end{tabular}
		\caption{Variation in CaII absorption line strength with redshift}
		\label{CaII evolution}
	\end{table*}
	
	\section{Conclusion}
	\label{conclusion}
	
	In this paper, we have presented a composite spectrum for Seyfert type 1
	galaxies. This spectrum has a sufficiently high signal to noise ratio
	to be useful as a template for identifying possible Seyfert 1
	candidates in future spectroscopic surveys of galaxies. It may also
	be used to improve k-corrections in flux measurements of Seyfert
	galaxies at different redshifts. For this purpose, users may wish to
	use the binned spectra shown in Figure~\ref{complete redshift
		evolution}, which are made available in the online version of this paper. The composite Seyfert 1 spectrum is also remarkably similar to the
	composite quasar spectrum published by \citet{VandenBerk2001}. This is
	consistent with quasars and Seyfert 1 being the same object (Type I
	AGN), as predicted by the unification model of AGN.
	
	\section*{Acknowledgements}
	
	We thank the anonymous referee whose insightful comments improved both the content and presentation of this paper. YW acknowledges IUCAA for hosting him on his sabbatical where a part
	of this work was completed. Specview is a  product of the Space Telescope Science Institute, which is operated by AURA for NASA.
	
	
	
	
	
	\bibliographystyle{mnras.bst}
	\bibliography{bibliography.bib}
	
	
	
	
	\bsp	
	\label{lastpage}
\end{document}